\documentclass[prb,aps,twocolumn,showpacs,groupedaddress,floatfix,amsmath,10pt]{revtex4}
\def\bc{\begin{center}}
\def\ec{\end{center}}
\def\beq{\begin{equation}}
\def\eeq{\end{equation}}
\def\bw{\begin{widetext}}
\def\ew{\end{widetext}}
\def\bea{\begin{eqnarray}}
\def\eea{\end{eqnarray}}

\renewcommand{\vec}[1]{\mbox{\boldmath$#1$}}
\usepackage{amssymb}
\usepackage{graphicx}
\usepackage{dcolumn}
\usepackage{bm}
\usepackage{subfigure}
\usepackage{verbatim}

\begin{document}

\title{Static and Dynamic Properties of Type-II Composite Fermion Wigner Crystals}
\author{Alex Archer and Jainendra K. Jain}
\affiliation{Department of Physics, Pennsylvania State University, University Park, PA 16802}

\date{\today}

\begin{abstract}
The Wigner crystal of composite fermions is a strongly correlated state of complex emergent particles, and therefore its unambiguous detection would be of significant importance. Recent observation of optical resonances in the vicinity of filling factor $\nu=1/3$ has been interpreted as evidence for a pinned Wigner crystal of composite fermions [Zhu {\em et al.}, Phys. Rev. Lett. {\bf 105}, 126803 (2010)].  We evaluate in a microscopic theory the shear modulus and the magnetophonon and magnetoplasmon dispersions of the composite fermion Wigner crystal in the vicinity of filling factors 1/3, 2/5, and 3/7. We determine the region of stability of the crystal phase, and also relate the frequency of its pinning mode to that of the corresponding electron crystal near integer fillings. These results are in good semiquantitative agreement with experiment, and therefore support the identification of the optical resonance as the pinning mode of the composite fermions Wigner crystal. Our calculations also bring out certain puzzling features, such as a relatively small melting temperature for the composite fermion Wigner crystal, and also suggest a higher asymmetry between Wigner crystals of composite fermion particles and holes than that observed experimentally.
\end{abstract}

\maketitle

\section{Introduction}

The fractional quantum Hall (FQHE) was discovered\cite{Tsui} when experimenters were searching for the Wigner crystal\cite{Wigner} (WC). The expectation was that once the kinetic energy is quenched by forcing all electrons into the lowest Landau level (LL), the interaction energy would govern the physics and induce a WC.\cite{Soviet} However, it turns out that, in a large range of filling factors of the lowest LL, the interaction energy favors the formation of composite fermions (CFs) instead, which form a liquid and produce the FQHE.\cite{Jain} However, when the interparticle separation is large compared to the size of a localized particle, namely the magnetic length $l$, the WC should be stabilized. This condition can be achieved in two ways: (i) by reducing the magnetic length, which can be  attained by going to very high fields (or low filling factors); and (ii) by increasing the interparticle separation, which can be  conveniently accomplished by considering systems close to integer fillings, where, to the zeroth order approximation, only the electrons or holes in the topmost partially filled LL are relevant. It is useful to differentiate between these two kinds of WCs, both because they originate from different interparticle interactions, and because the density of the relevant particles in the latter actually depends on the magnetic field, vanishing at the magnetic field where the filling is an integer.  We will therefore label these crystals ``type-I" and ``type-II" Wigner crystals, respectively.  Much experimental work has been performed to investigate such crystalline states.\cite{Chen,Chen2,Lewis,Zhu1,Samban} Recently, microwave resonance experiments\cite{Zhu1} have extensively probed regions at low fillings and also near integer fillings and revealed resonances in the conductivity that are interpreted as the pinning mode of a WC. 

Interestingly, even though the formation of composite fermions was responsible for inhibiting the {\em electron} WC in a broad range of filling factors, composite fermions themselves can form a WC, which is called a composite fermion WC (CFWC). Composite fermions\cite{Jain} are bound states of electrons and quantized vortices. They experience an effective magnetic field $B^*$ and form Landau-like levels called $\Lambda$ levels ($\Lambda$Ls). Their filling factor $\nu^*$ is related to the electron filling factor by $\nu=\nu^*/(2p\nu^*\pm 1)$, where $2p$ is the number of vortices attached to composite fermions. As for electrons, there exist two kinds of CFWCs:

{\em Type-I CFWC}: It has been shown theoretically that at very low fillings, the CFWC has lower energy than the electron WC, i.e. the actual ground state is at type-I CFWC.\cite{Chang,Yi,Nare,Jeon,QD2} Essentially, electrons capture fewer than the maximum number of vortices available to them, and use the remaining degrees of freedom to form a WC. The CFWC has been shown to be an excellent approximation of the exact state in numerical diagonalization studies.\cite{Chang} These studies predict that composite fermions with a large number of attached vortices can occur in type-I CFWC, and that there should be transitions, as a function of the filling factor, from WC of one flavor of composite fermions to another.

{\em Type-II CFWC}: The CFWC can also occur at CF fillings close to $\nu^*=n$, which correspond to electron fillings close to $\nu=n/(2pn\pm 1)$.  The state at $\nu^*=n$ contains an integer number of fully occupied $\Lambda$Ls, and at nearby fillings the system has either a small density of composite fermions in an otherwise empty $\Lambda$L or a small density of CF holes in an otherwise full $\Lambda$L. These will form a type-II CFWC.\cite{bLee} 

We compute in this paper the static and dynamic properties of the type-II CFWC. The primary motivation of our work comes from the remarkable recent experiments by Zhu {\em et al.}\cite{Zhu1} where  a resonance was observed in a narrow region around $\nu=1/3$ and interpreted as the pinning mode of the type-II CFWC. A comparison of the results of our calculation with the experimental observations shows a good agreement, thus supporting the formation of a CFWC. 

We study the CFWC by making an exact mapping of the problem of composite fermions in a partially filled $\Lambda$L to that of fermions in the lowest LL interacting via an effective interaction. Various quantities can then be computed using the familiar methods developed in Refs. [\onlinecite{BM,MZ,bLee,Ettouhami}]. (The type-II CFWC was also studied by Scarola {\em et al.}\cite{aLee,bLee} in a variational study that compared the energies of the WC, bubble crystal, or stripe phases of composite fermions. That work, however, did not compute the shear modulus of the CFWC or the magnetophonon and magnetoplasmon dispersions, and also did not focus on the phase diagram of the CFWC.) Our main conclusions are as follows. We estimate the ranges of stability for the CFWC around the filling factors $\nu=\frac{1}{3},\frac{2}{5},\frac{3}{7}$. The theoretical range of stability around $\nu=\frac{1}{3}$ is consistent with the interpretation of microwave resonance experimental results around $\frac{1}{3}$ as a signature of the CFWC.\cite{Zhu1} We predict CFWC also around $\frac{2}{5}$ and $\frac{3}{7}$, except in much narrower ranges of filling factor. We present results for the dispersions of the phonons as well as magnetoplasmon of the CFWC. Using a model of Chitra et. al.\cite{Chitra} we argue that the frequency of the pinning resonance of the CFWC is expected to be of similar magnitude as that of the corresponding electron WC, and also determine the magnetic field dependence; these are also consistent with experiments. 

A curious observation is that the Hartree Fock treatment is much more successful for the CFWC than for the electron WC. For example, Maki and Zotos\cite{MZ} found, with similar methods as those employed below, that the electron WC in the lowest LL is stable in the range $\nu<0.45$ (and $1>\nu>0.55$ by particle hole symmetry in the lowest LL), which is grossly inconsistent with experiments that demonstrate WC only below approximately $\nu<1/7$. On the other hand, the stability range we find for the CFWC is quite consistent with experiment. While this difference may seem surprising, it is easily understood. The Hartree-Fock approach fails for electrons in the lowest LL because they form strongly correlated states of composite fermions, not captured by the Hartree Fock approach. In contrast, the interaction between composite fermions is such that the CFWC state is destabilized before composite fermions can form their own strongly correlated FQHE states.  For example, the $4/11$ FQHE state, which is a {\em fractional} QHE of composite fermions\cite{Pan03,Chang2,WYQ,Goerbig} at the composite fermion filling factor of $\nu^*=1+1/3$, is outside the CFWC region, which persists, according to our calculation, only up to $\nu^*\approx 1.10$.

The remainder of the paper is organized as follows. We describe the mapping from composite fermions in a partially filled $\Lambda$L to fermions in the LLL. Next, we solve the equations of motion for the CFWC, and calculate the shear modulus and the dispersion relations for the magnetophonon and magnetoplasmon collective modes of the this crystal. Finally, we conclude with a discussion of how our results compare with experimental observations.

\section{Model}

Electrons in the lowest Landau level exhibit exotic properties, which result from the formation of composite fermions, bound states of electrons and an even number of topological vortices. 
At the rational filling factors $\nu=\frac{n}{2np+1}$ the ground state contains precisely $n$ filled $\Lambda$Ls of composite fermions. Of interest in this paper are filling factors close to these rational values, when the CF filling factor is given by 
\begin{equation}
\nu^*=n\pm \bar{\nu}^*
\end{equation}
In what follows below, we assume that the magnetic field is sufficiently strong that the system is fully spin polarized and LL mixing is negligible. Our program proceeds along the following steps:

(i) We treat the $n$ filled $\Lambda$Ls as inert, and formulate the problem in terms of either solely the composite fermions in the $(n+1)^{th}$ $\Lambda$L (for $\nu^*=n+\bar{\nu}^*$) or the missing composite fermions in the $n^{th}$ $\Lambda$L (for $\nu^*=n-\bar{\nu}^*$). The physics that we wish to investigate lies fully in the dynamics of these composite fermions.  These are sometimes called CF quasiparticles and CF quasiholes, but we will refer to them simply as composite fermions below, while remembering that their filling factor is $\bar{\nu}^*$, and that their density vanishes at $\nu= \frac{n}{2np+1}$ (where $\bar{\nu}^*= 0$).   We do not include the physics of $\Lambda$L mixing in our work; it will modify the form of the inter-CF interaction, but we do not believe that would significantly alter the results.

(ii) The two-body Coulomb interaction between electrons generates a complex interaction between composite fermions that contains two, three and higher body terms. We assume that the physics is dominated by the two-body interaction (determined in the next section). This should be a good approximation at sufficiently small densities of the composite fermions; because the CF filling is found to be small in the entire range of stability of the CFWC determined below, we believe that the approximation of the neglect of three and higher body interaction terms in the CF Hamiltonian does not cause significant corrections. There is also numerical evidence that this is a good approximation\cite{WQ}, but a systematic study of the importance of higher order interactions between particles as a function of system size has not been undertaken.\cite{bLee}

(iii) We define the equivalent problem of fermions at filling factor $\bar{\nu}^*$ in the lowest LL interacting via an appropriate effective interaction.  The effective interaction used here has been shown to reproduce the desired interaction pseudopotentials with extremely good accuracy.

(iv) We calculate the properties of the WC of fermions at $\bar{\nu}^*$ in a standard Hartree Fock approach.\cite{MZ,Ettouhami}  In particular, the instability of the CFWC is signaled by the shear modulus becoming negative. 

(v) Disorder is neglected throughout. This makes our results unreliable at very small values of $\bar{\nu}^*$, where the distance between the (relevant) composite fermions is so large that disorder will likely dominate over interaction, producing a localized or a glassy phase. However, we expect the neglect of disorder to be a good first approximation for weak disorder and not too large a lattice spacings; here the WC order ought to persist meaningfully over several lattice spacings.

\begin{figure}[h]
\includegraphics[width=0.45\textwidth]{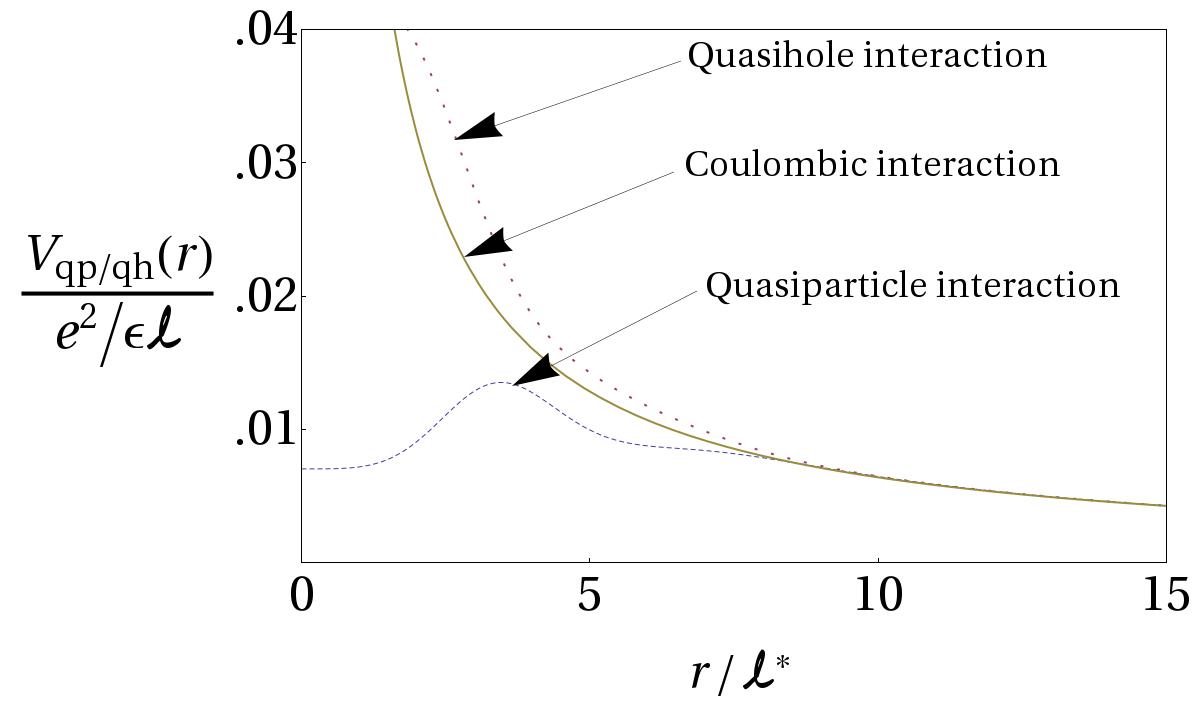}
\caption{Plot of the semiclassical interaction between composite fermions near $\nu^*=1/3$, both for composite fermions in the second $\Lambda$ level (labeled ``quasiparticle") and CF quasiholes in the lowest $\Lambda$ level (``quasihole"). At long distances, both interactions are Coulombic, but have more complicated forms at short distance.
}\label{effpot}
\end{figure}

\subsection{Two-body Interaction between composite fermions}

The CF wave functions can be obtained from the electron wave functions by the standard mapping:
\begin{equation}\label{CFwf}
 \Psi^{\alpha}_{\frac{\nu^*}{2p\nu^*\pm 1}}=P_{\rm LLL}\prod_{j<k}(z_j-z_k)^{2p}\Phi^{\alpha}_{\pm\nu^*}
\end{equation}
Here $\alpha$ is a quantum number labeling the state, $\nu^*$ is the CF filling factor, and $P_{\rm LLL}$ is the LLL projection operator. The interaction between two composite fermions in the $(n+1)^{\rm th}$ $\Lambda$L is determined by computing the CF pseudopotentials,\cite{Quinn0,Park,aLee,bLee,QD2} i.e. the energies of the pair in definite relative angular momentum states, by analogy to the electron pseudopotentials in the $n^{\rm th}$ LL. The pseudopotential for two electrons in relative angular momentum $L$ is defined as\cite{Haldane}:
\begin{equation}
V^{n}_L=<\!\Phi_{L}^{n,++}|\frac{e^2}{\epsilon l}\frac{1}{|\vec{r}_1-\vec{r}_2|}|\Phi_{L}^{n,++}\!>
\label{eps}
\end{equation}
where $\Phi_{L}^{n,++}$ is the wave function of two electrons in the $n^{\rm th}$ LL in relative angular momentum $L$.  The CF pseudopotentials are analogously defined as
\begin{equation}
V^{\rm CF,n}_L=<\!\Psi_{{\rm CF},L}^{n,++}|\sum_{i<j}\frac{e^2}{\epsilon l}\frac{1}{|\vec{r}_i-\vec{r}_j|}|\Psi_{{\rm CF},L}^{n,++}\!>
\end{equation}
where 
\begin{equation}
\Psi_{{\rm CF},L}^{n,++}=P_{\rm LLL}\prod_{j<k}(z_j-z_k)^{2p}\Phi_{L}^{n,++}
\end{equation}
While for the calculation of of the electron pseudopotentials $V^{n}_L$, it is sufficient to consider only two electrons in the $n^{\rm th}$ LL, the evaluation of the CF pseudopotentials requires a full many-particle calculation, because the Jastrow factor correlates all composite fermions with one another (including those in lower $\Lambda$ levels). Nonetheless, using the standard methods\cite{JKam}, the relevant integrals can be evaluated by the Metropolis Monte Carlo method, and the thermodynamic limits for the CF pseudopotentials can be obtained. We refer the reader to the literature for further details.\cite{Quinn0,Park,aLee,bLee,QD2}

We next define a problem of fermions confined to the lowest LL that have an effective interaction $V^{\rm eff}(r)$ that produces the above pseudopotentials:
\begin{equation}
 <\!\Phi_{L}^{0,++}|V^{\rm eff}(\vec{r})|\Phi_{L}^{0,++}\!>=V^{\rm CF,n}_L
\end{equation}
where the superscript $0$ refers to the lowest LL. There are many possible choices for $V^{\rm eff}(\vec{r})$ which produce the same effective interaction.\cite{commpseudo} We find it convenient to use the effective interaction given by Lee, Scarola, and Jain\cite{aLee,bLee}: 

\begin{equation}\label{effint}
 V^{\rm eff}(r^*)=\frac{e^2}{l}\left[\sum_i\left( c_ir^{*2i} e^{-r^{*2}}\right) + \frac{(2n+1)^{-5/2}}{r^*}\right]
\end{equation}
The asterisk on $r^*$ indicates that the distance is being measured in units of the effective magnetic length $l^*=\sqrt{\hbar c/e |B^*|}$. The last term captures the long distance limit of the interaction between composite fermions: $(e^*)^2/r=(\frac{e}{2n+1})^2/{r^*l^*}=(\frac{e}{2n+1})^2/{r^*}(2n+1)^{1/2}l$, which is the Coulomb interaction between particles of fractional charge $e^*=e/(2n+1)$ at a distance $r$.\cite{CFtext} The number of terms kept in the sum, and the coefficients $c_i$, are determined to produce a good approximation for the pseudopotentials obtained from the microscopic theory.  We use below the $c_i$ given in Ref.[\onlinecite{bLee}].

In what follows, we will use $l^*$ as the unit of length and $e^2/l$ as the unit of energy. We will omit the asterisk from the lengths for notational convenience. We will further set $\hbar=1$ and $c=1$. In certain places of relevance, the units will be explicitly restored.

\subsection{Semiclassical model for CFWC}
Now, we consider the dynamics of fermions in the LLL interacting with an effective interaction. Since we are dealing with fermions in the LLL, they all have the same kinetic energy. Therefore, the ground state of the system is decided entirely by interactions between the fermions. Since we will be working at low filling factors, we make the ansatz that our system is described by a crystal state\cite{Wigner,Soviet}, for which we choose the hexagonal (or triangular) lattice symmetry, which it is known to give the lowest energy for the 2D classical WC.\cite{BM}   The localized single particle states are wave packets:
\begin{equation}
\psi_{\vec{{\scriptstyle R}}}(\vec{r})=\frac{1}{(2\pi)^{1/2}}e^{-\frac{1}{4}((\vec{{\scriptstyle r}}-\vec{{\scriptstyle R}})^2-2i(\vec{{\scriptstyle r}}\times \vec{{\scriptstyle R}})\cdot \hat{z})}
\end{equation}
centered at hexagonal lattice sites 
\begin{equation}
\vec{R}=a\left(n+{1\over 2}m,{\sqrt{3}\over 2}m\right),
\label{R}
\end{equation} 
where $n,m$ are integers and $a$ the lattice constant.
The Maki-Zotos\cite{MZ} wave function is obtained by placing a Gaussian wave packet at each lattice site and then antisymmetrizing:
\begin{equation}
\Psi_{\{\vec{{\scriptstyle R_j}}\}}(\{\vec{r_i}\})=\det(\psi_{\vec{{\scriptstyle R_i}}}(\vec{ r}_j))
\end{equation}

The problem can be recast as a semiclassical crystal dynamics problem by calculating the expectation value of the effective interaction:
\begin{eqnarray}\label{VMZ}
&&\frac{<\! \Psi_{\{\vec{{\scriptstyle R_j}}\}}(\{\vec{r_i}\}) |\sum_{i<j}V^{\rm eff}(r_{ij})| \Psi_{\{\vec{{\scriptstyle R_j}}\}}(\{\vec{r_i}\}) \!>}{<\! \Psi_{\{\vec{{\scriptstyle R_j}}\}}(\{\vec{r_i}\}) |\Psi_{\{\vec{{\scriptstyle R_j}}\}}(\{\vec{r_i}\}) \!>}=\nonumber\\
&&\hspace{1cm}\sum_{i<j}V(R_{ij}) +\sum_{i,j,k}V_3(R_i,R_j,R_k)+\ldots
\end{eqnarray}
where the sums on the right are infinite sums over hexagonal lattice sites, $R_{ij}=|\vec{R}_i-\vec{R}_j|$, and $V\!(R_{ij})$ is given by:
\begin{equation}\label{V2}
V\!(R_{ij})=\frac{\int r\;dr\;V^{\rm eff}\!(r)e^{-r^2/4}(I_0(\frac{1}{2}R_{ij}r)-J_0(\frac{1}{2}R_{ij}r))}{4\sinh(R^{2}_{ij}/4)}
\end{equation}
The derivation of these expressions closely follows the work of Maki and Zotos\cite{MZ}, and is given in Appendix A. As stated above, we neglect $n\!\geq\!3$ body terms in Eq. (\ref{VMZ}). 

The individual two body terms are interpreted as semiclassical energy of particles located at the lattice sites of a hexagonal lattice. We have calculated the energy for both composite fermions around each of the filling factors $\nu=\frac{1}{3},\frac{2}{5},\frac{3}{7}$ using the effective real space interactions discussed above. With the particular form of the effective interaction used (see Eq.~(\ref{effint}))\cite{bLee}, the integrals in Eq.~(\ref{V2}) can be evaluated analytically using Mathematica, but the resulting expression is too long to reproduce here. In Fig.~\ref{effpot} we plot the semiclassical interaction given by Eq.~(\ref{V2}) using the values of $c_i$ for composite fermions around $\nu=1/3$ given by [\onlinecite{bLee}]. At short range, these interactions have a complicated form as a function of the distance, and can even be attractive, but at long distance the interaction is always repulsive. The attractive portion of the interaction between the composite fermions might seem to suggest that the crystal is unstable to pairing, but the issue is somewhat subtle because the gain in energy due to pairing is accompanied by a competing enhancement in the interaction energy between the pairs. It has been shown by an explicit calculation reported in Ref.~[\onlinecite{bLee}] that for sufficiently small values of $\bar{\nu}^*$ a WC state is energetically preferred over the paired liquid state, although the formation of bubble crystals, where each lattice site supports more than one composite fermion, can result from an attractive interaction in some filling factor ranges.  We will not consider the possibility of bubble crystals below, and assume that there is a single CF at each lattice site.

\section{Equations of Motion and Shear modulus}

Next we calculate the lattice dynamics for a crystal of composite fermions in the harmonic approximation.\cite{LDHM}
We interpret the sum $\sum_{i<j}V(R_{ij})$ as the zero-point energy of the CFWC with composite fermions located at the lattice sites of a hexagonal lattice.  We perturb the particle's equilibrium location $\vec{R}_i$ by a small, time dependent quantity $\vec{u}_i(t)$:
\begin{equation}\vec{R}_i \rightarrow \vec{R}_i+\vec{u}_i\end{equation}
We expand $V\!(|\vec{R}_i+\vec{u}_i-\vec{R}_j-\vec{u}_j|)$ to quadratic order in $\vec{u}_i,\vec{u}_j$ and write the equation of motion for the $i^{th}$ particle:
\begin{widetext}
\begin{equation}\label{xmot}
m^*\ddot{u}_i^x=eB^*\dot{u}_i^y-\sum_j  {1\over l^{*2}} \left[\frac{\partial^2V(R_{ij})}{\partial R_{ij,x}^2}(u_i^x-u_j^x)+
\frac{\partial^2V(R_{ij})}{\partial R_{ij,x} \partial R_{ij,y}}(u_i^y-u_j^y)\right] 
\end{equation}
\begin{equation}\label{ymot}
m^*\ddot{u}_i^y=-eB^*\dot{u}_i^x-\sum_j {1\over l^{*2}} \left[\frac{\partial^2V(R_{ij})}{\partial R_{ij,y}^2}(u_i^y-u_j^y)+\frac{\partial^2V(R_{ij})}{\partial R_{ij,x} \partial R_{ij,y}}(u_i^x-u_j^x)\right] 
\end{equation}
\end{widetext}
In writing these equations, we have assumed that the dynamics of composite fermions\cite{Evers} corresponds to particles of mass $m^*$ in an effective magnetic field $B^*=B/(2n+1)$. The CF mass $m^*$ is unrelated to the electron mass.\cite{CFtext} The factor of $(l^*)^{-2}=eB^*$ results from expressing the lengths in units of $l^*$, while recognizing that the argument of $V(R_{ij})$ is already in units of $l^*$. This is consistent with the expectation that the magnetic field renormalization from $B$ to $B^*$ upon the formation of composite fermions is associated with an analogous electric field renormalization.\cite{Goldhaber}
Assuming a solution of the form $\vec{u}_i(t)=\vec{u}_{\vec{\scriptstyle{k}}}e^{-i\omega t-i\vec{\scriptstyle{k}}\cdot \vec{\scriptstyle{R}}_i}$ the equations of motion in the $x$ and $y$ directions take the form:
\begin{equation}\label{xeq}
 \omega^2u_{\vec{\scriptstyle{k}}}^x=\left[i\omega\omega^*_c+\omega^*_c\Phi_{xy}\!(\vec{k})\right]u_{\vec{\scriptstyle{k}}}^y+\omega^*_c\Phi_{xx}\!(\vec{k})u_{\vec{\scriptstyle{k}}}^x
\end{equation}
\begin{equation}\label{yeq}
 \omega^2u_{\vec{\scriptstyle{k}}}^y=\left[-i\omega\omega^*_c+\omega^*_c\Phi_{xy}\!(\vec{k})\right]u_{\vec{\scriptstyle{k}}}^x+\omega^*_c\Phi_{yy}\!(\vec{k})u_{\vec{\scriptstyle{k}}}^y
\end{equation}
where 
\begin{equation}
\omega^*_c=\frac{eB^*}{m^*}={eB\over (2n+1) m^*}, 
\end{equation}
and $\Phi_{\alpha\beta}$ is given by
\begin{equation}\label{phiab}
\Phi_{\alpha\beta}\!(\vec{k})=\sum_j\frac{\partial^2V\!(R_j)}{\partial{R_{\alpha}R_{\beta}}}(1-\cos({\vec{k}\!\cdot\!\vec{R}_j})).
\end{equation}
Here, $\alpha$ and $\beta$ denote $x$ and $y$ components.

Equations (\ref{xeq}) and (\ref{yeq}) can be diagonalized to obtain the dispersion $\omega^2_{\pm}\!\left(\vec{k}\right)$:
\begin{widetext}
\begin{equation}\label{disp}
\omega^2_{\pm}(\vec{k})=\frac{1}{2}\left(\omega^{*2}_c+\omega^*_c\left[\Phi_{xx}(\vec{k})+\Phi_{yy}(\vec{k})\right]\right)\pm\sqrt{\frac{1}{4}(\omega^{*2}_c+\omega^*_c[\Phi_{xx}(\vec{k})+\Phi_{yy}(\vec{k})])^2+\omega^{*2}_c[\Phi_{xy}^2(\vec{k})-\Phi_{xx}(\vec{k})\Phi_{yy}(\vec{k})]}
\end{equation}
\end{widetext}

To ascertain the regions of filling factor where the CF crystal is stable, we need to examine the phonon dispersion in the long wavelength limit, $k\rightarrow0$. In the limit of small $k$ we can write $\Phi_{\alpha\beta}$ as\cite{MZ,Ettouhami}:
\begin{equation}\label{lphiab}
\Phi_{\alpha\beta}=\left(\frac{\bar{\nu}^*}{k}+(C_L-C_t)\right)k_{\alpha}k_{\beta}+\delta_{\alpha\beta}C_t k^2
\end{equation}
where $C_t$ is the shear modulus and $C_L$ the coefficient in the longitudinal mode. Both have units of energy. In two dimensions, it can be shown that the dynamical matrix of a two dimensional hexagonal crystal with central forces 
characterized by two parameters, here conveniently chosen to be $C_t$ and $C_L$.\cite{Ettouhami} As shown in Ref. \onlinecite{MZ}, the classical value of $C_t$ for particles with charge $e^*$ in a magnetic field $B^*$ and filling factor $\bar{\nu}^*$ 
is given by $C_{t_0}=0.09775 ({\bar{\nu}^*})^{1/2} e^{*2}/l^*$, which, when converted into our energy units of $e^2/l$, reduces to $C_{t_0}=0.09775(\bar{\nu}^*)^{1/2}/(2n+1)^{5/2}$. The parameter $C_{L_0}$ is given by $C_{L_0}=-5C_{t_0}$.\cite{MZ} Non-classical properties of the CFWC can be captured by measuring the variation of $C_t$ and $C_L$ with respect to their classical counterparts in the region where the crystal is stable. Substituting into Eq. (\ref{disp}) produces:
\begin{widetext}
\begin{equation}
\omega^2_{\pm}(\vec{k})=\frac{1}{2}(\omega^{*2}_c+\omega^*_c(\frac{\bar{\nu}^*}{k}+C_L+C_t)k^2)\pm\sqrt{\frac{1}{4}(\omega^{*2}_c+\omega^*_c(\frac{\bar{\nu}^*}{k}+C_L+C_t)k^2)^2-\omega^{*2}_cC_t(\frac{\bar{\nu}^*}{k}+C_L)k^4}
\end{equation}
\end{widetext}
where we have kept terms up to quadratic order in $k$. Expanding in powers of $k$, we get:
\begin{equation}
\omega^2_{+}(\vec{k}) =(\omega^*_c)^2+O(k), 
\end{equation}
\begin{equation}
\omega^2_{-}(\vec{k})=C_t\bar{\nu}^* k^3+O(k^4).
\end{equation}
The $\omega_{+}^2(k)$ mode is called the magnetoplasmon mode and the $\omega_{-}^2(k)$ mode is called the magnetophonon mode. 
For negative values of $C_t$ the magnetophonon frequency becomes imaginary in the long wave length limit, indicating an instability of the CF crystal.

\begin{figure*}[h!]
\includegraphics[width=.9\textwidth]{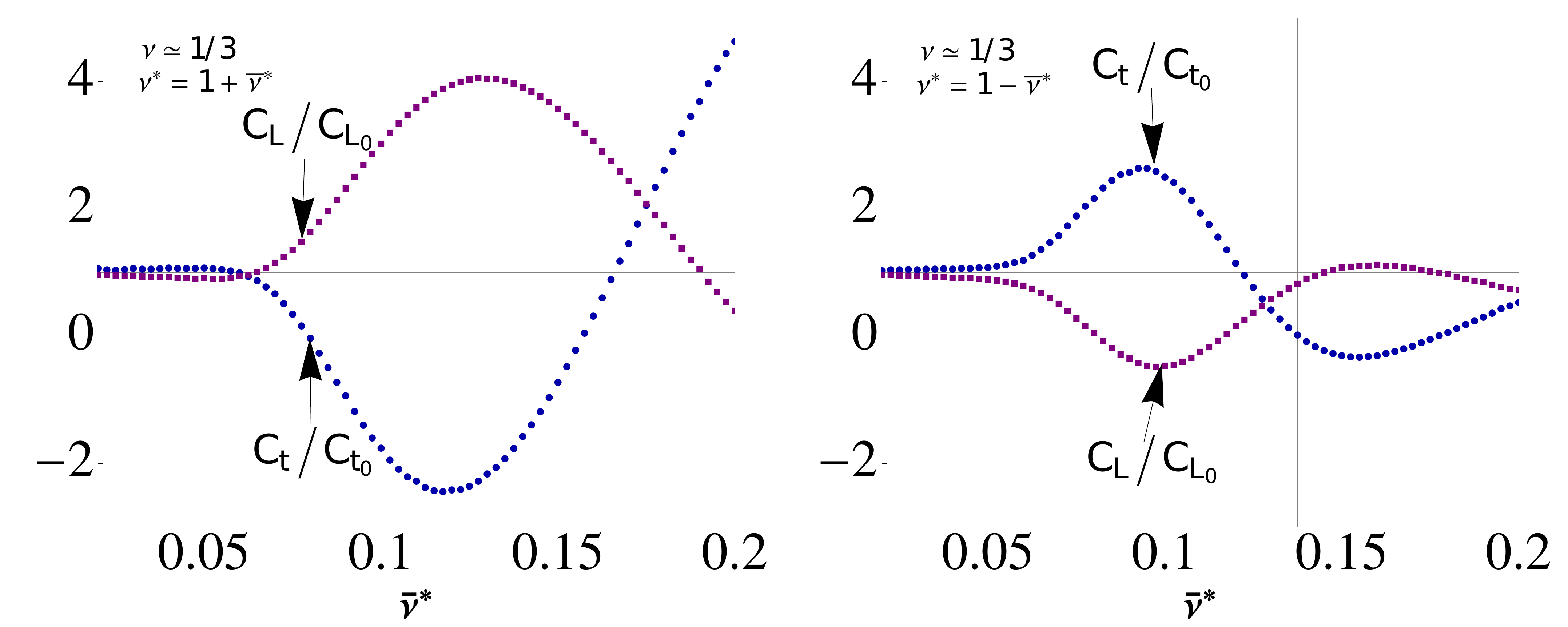}
\includegraphics[width=.9\textwidth]{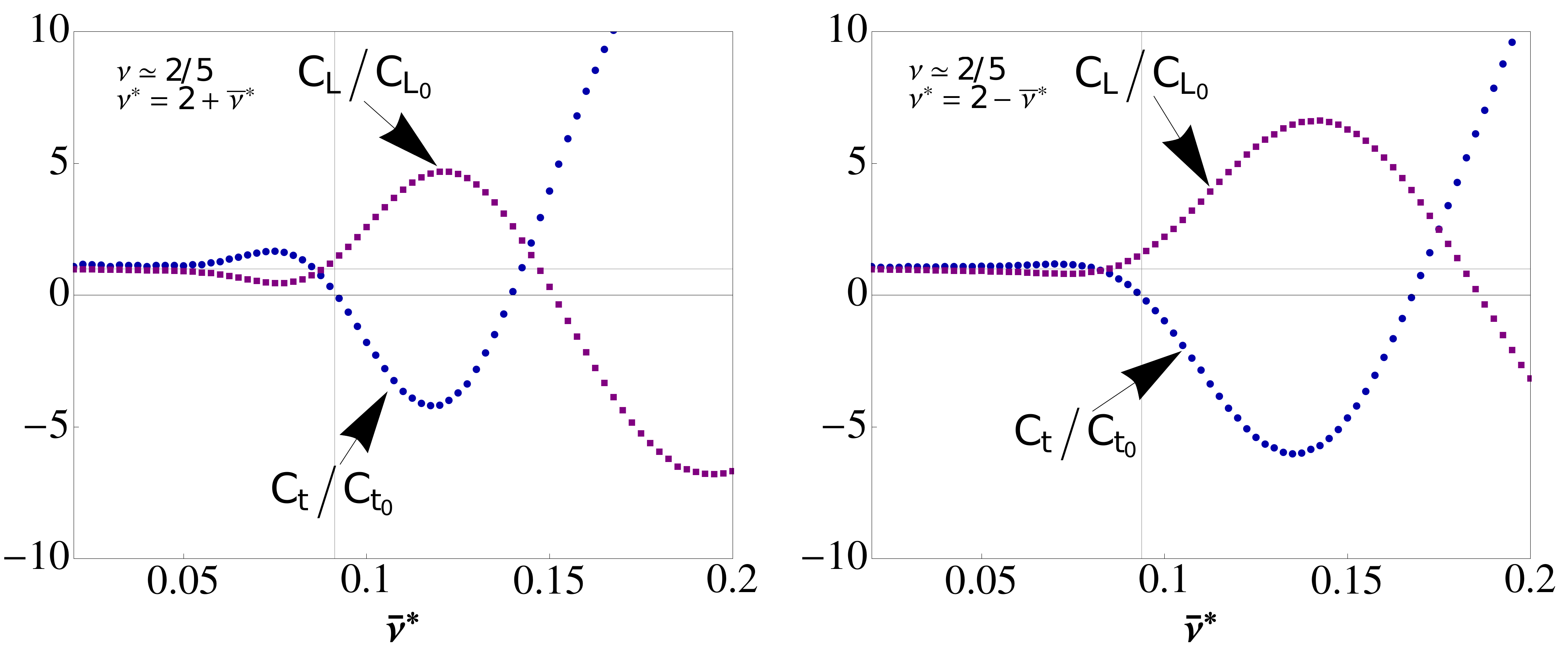}
\includegraphics[width=.9\textwidth]{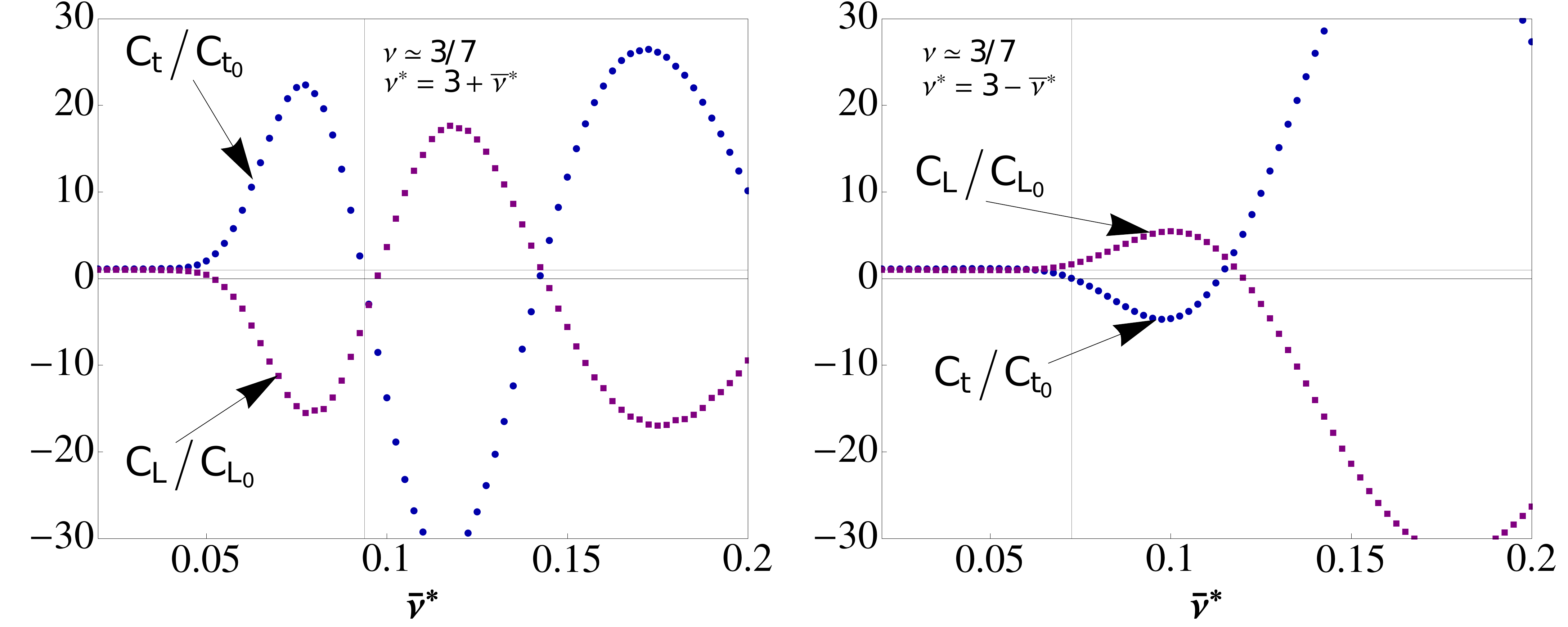}
\caption{Plot of the shear modulus $C_t$ and the coefficient in the longitudinal mode $C_L$ for the CFWC at $\nu^*=n+\bar{\nu}^*$ (left panels) and $\nu^*=n-\bar{\nu}^*$ (right panels), as a function of  the partial filling factor $\bar{\nu}^*=|\nu^*-n|$. The top row corresponds to $\nu^*\approx 1$ ($\nu\approx 1/3$), the middle row to $\nu^*\approx 2$ ($\nu\approx 2/5$), and the bottom row to $\nu^*\approx 3$ ($\nu\approx 3/7$).  $C_t$ and $C_L$ are normalized by their classical counterparts $C_{t_0}\text{ and }C_{L_0}$. The vertical lines indicate the filling factor where the shear modulus becomes negative and the CFWC becomes unstable.}\label{c13qp}
\end{figure*}

\begin{figure}[h]
\begin{center}
\includegraphics[width=0.45\textwidth]{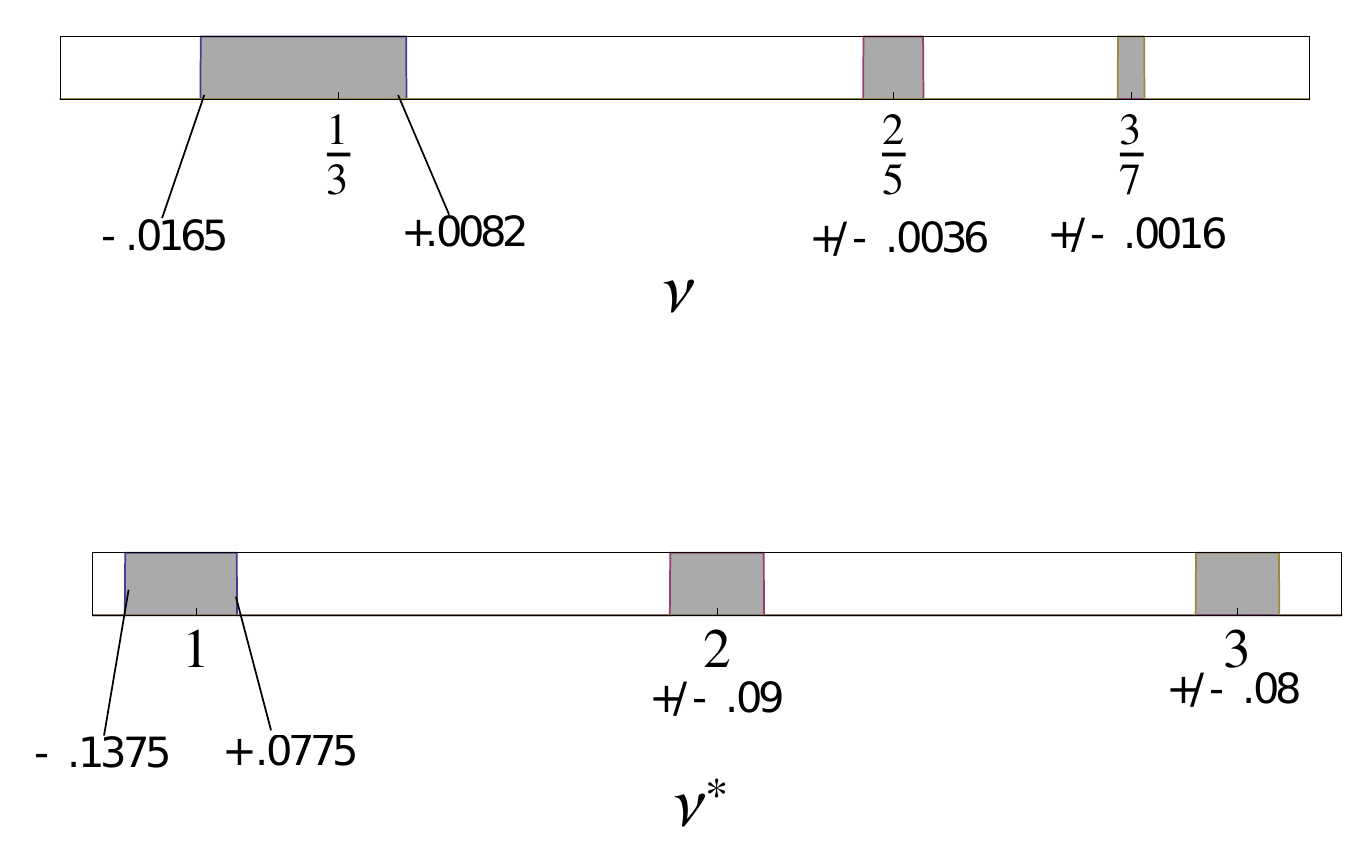}
\caption{Phase diagram for the CFWC. The shaded regions show the theoretically calculated stability regions of the CFWC in terms of the real filling factor $\nu$ (top) and the CF filling factor $\nu^*$ (bottom). 
}\label{stabreg}
\end{center}
\end{figure}

\section{Results and discussion}

We give in this section the results of our calculation, and, where possible, compare them to experiments. 

\subsection{Shear modulus and the region of stability}

To calculate the shear modulus, we evaluate the dynamical matrix $\Phi_{\alpha\beta}$ of Eq. \!\!(\ref{phiab}) numerically for several different values of $k$ and fit the result to the small $k$ limit form of Eq. (\ref{lphiab}). Further details are given in Appendix B. In Fig. \ref{c13qp} we show our results for the CFWC moduli around $\nu=1/3,2/5,3/7$. In this figure, we plot $C_t$ and $C_L$ normalized with respect to the classical shear modulus of particles with fractional charge as a function of $\bar{\nu}^*$.  Notice that as $\bar{\nu}^*$ approaches zero, all of the crystal moduli become equal to the classical crystal moduli. We find that at sufficiently low filling factors (approximately $\bar{\nu}^*<1/12$), the CF moduli are indistinguishable from their classical crystal counterparts. The non-classical behavior of the CFWC becomes manifest only at somewhat larger values of $\bar{\nu}^*$ and, indeed, makes a significant correction to the phase boundary marked by the position where the shear modulus changes its sign.

The CFWC becomes unstable when $C_t$ becomes negative. We depict the stability regions around the filling factors $\frac{1}{3},\frac{2}{5},\frac{3}{7}$ in Fig. \ref{stabreg}, which is one of the principal results of our work. The predicted range near $\nu=1/3$ is nicely consistent with the experiment of Zhu {\em et al.}\cite{Zhu1}, who find a range $\sim$ $1/3 \pm .015$, determined by where a microwave resonance is resolved. This differs by an order magnitude from the range $1\pm .15$ for the electron crystal.\cite{Zhu1}  In fact, as also noted by Zhu {\em et al.}, when measured in terms of the CF filling, the range near 1/3 is given by $\nu^*=0.88-1.15$, which is more similar to the observed stability region of the $\nu=1$ WC. Overall, our calculations provide strong support to the identification of the microwave resonance experiments in terms of the CFWC. We also predict regions of stability for the CFWC around $\nu=2/5 \text{ and } 3/7$, which, however, are much narrower and may be harder to observe, especially considering that at very small $\bar{\nu}^*$ the disorder may dominate producing a localized glassy phase.

We note that we mark the region of stability as the filling $\bar{\nu}^*$ where the shear modulus {\em first} becomes negative as the $\bar{\nu}^*$ is increased from zero. According to our calculations the shear modulus becomes positive again at some larger fillings, to which we do not assign any physical significance -- we believe that those regions are outside the regime where our approximations are valid. It was noted in previous calculations\cite{bLee} that composite fermions can possibly form bubble crystal and stripe phases at larger values of $\bar{\nu}^*$.

\subsection{Magnetophonon and magnetoplasmon dispersions}

\begin{figure*}[h!]
\includegraphics[width=0.8\textwidth]{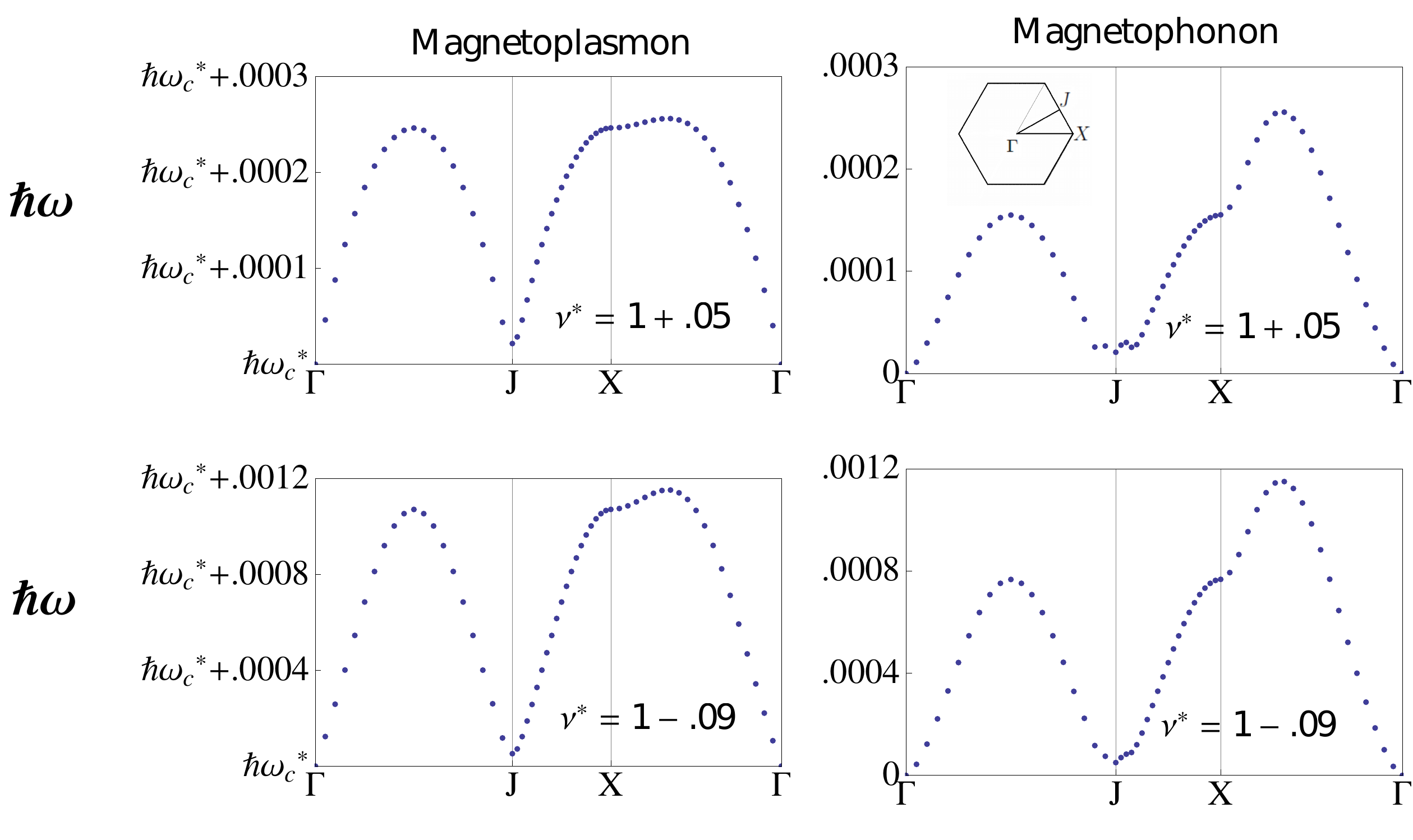}
\caption{Magnetoplasmon and magnetophonon dispersion of the CFWC near $\nu=1/3$ around the irreducible element of the first Brillouin zone. The cyclotron frequency is given by $\omega_c^*=eB^*/m^*$ where $B^*$ is the effective magnetic field and $m^*$ is the CF mass; the results are quoted in units of $e^2/\epsilon l$. For illustration, the dispersions are given at filling factors where the CF shear modulus has a maximum, i.e. where the CFWC is most robust, both below and above $\nu^*=1$. 
The irreducible element of the FBZ is shown in the upper right hand figure.}\label{13qp}
\end{figure*}

\begin{figure*}[h!]
\includegraphics[width=0.8\textwidth]{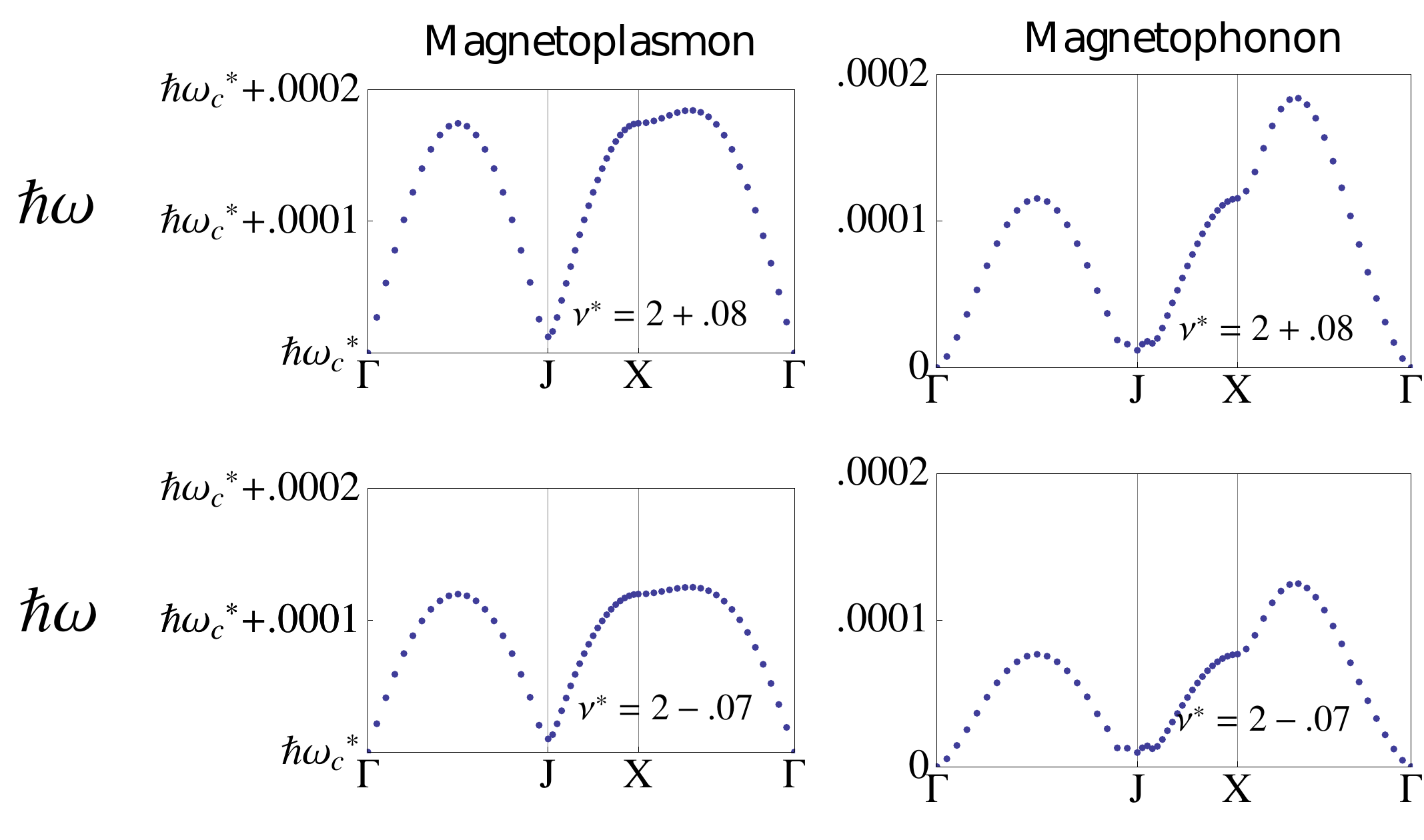}
\caption{Magnetoplasmon and magnetophonon dispersion of the CFWC near $\nu=2/5$ around the irreducible element of the first Brillouin zone. The cyclotron frequency is given by $\omega_c^*=eB^*/m^*$ where $B^*$ is the effective magnetic field and $m^*$ is the CF mass; the results are quoted in units of $e^2/\epsilon l$. For illustration, the dispersions are given at filling factors where the CF shear modulus has a maximum, i.e. where the CFWC is most robust, both below and above $\nu^*=2$. }\label{25qp}
\end{figure*}

\begin{figure*}[ht]
\includegraphics[width=0.8\textwidth]{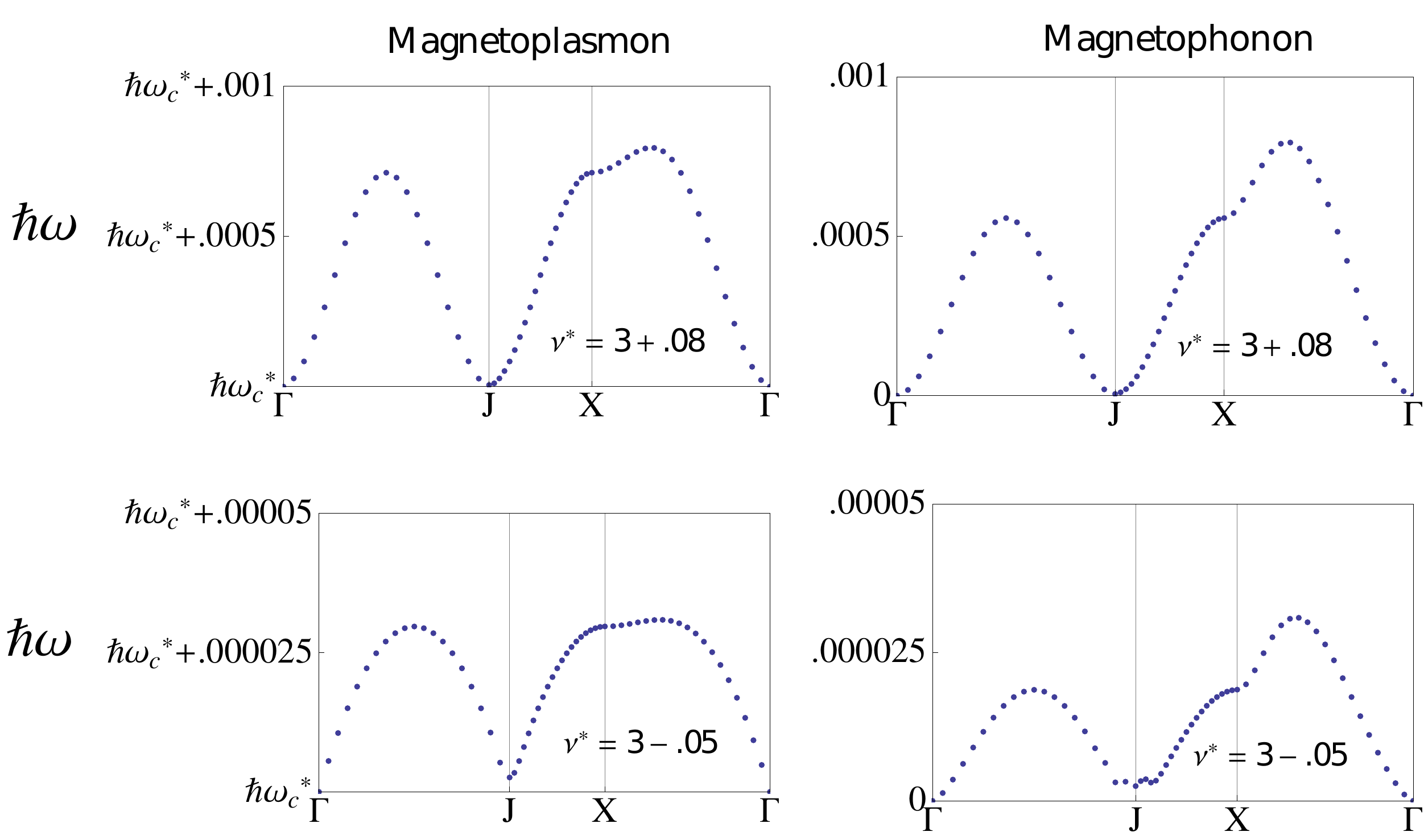}
\caption{Magnetoplasmon and magnetophonon dispersion of the CFWC near $\nu=3/7$ around the irreducible element of the first Brillouin zone. The cyclotron frequency is given by $\omega_c^*=eB^*/m^*$ where $B^*$ is the effective magnetic field and $m^*$ is the CF mass; the results are quoted in units of $e^2/\epsilon l$. For illustration, the dispersions are given at filling factors where the CF shear modulus has a maximum, i.e. where the CFWC is most robust, both below and above $\nu^*=3$. }\label{37qp}
\end{figure*}

In Fig. \ref{13qp}, we plot the dispersions of the magnetoplasmon and magnetophonon collective modes for the CFWC around the irreducible element of the first Brillouin zone (FBZ) around $\nu=1/3$. For small wave vectors, the dispersion has the form $\omega\propto k^{3/2}$, which is identical to the small $k$ behavior of the classical WC dispersion. The dispersion relations for the magnetophonons and magnetoplasmons for the CFWC around $\frac{2}{5}$ and $\frac{3}{7}$, shown in Figs. \ref{25qp} and \ref{37qp}, have nearly the same shape and appear to be scaled versions of the dispersion relations shown for the crystals around $\nu=1/3$. 

The magnetophonon and magnetoplasmon dispersions of the WC obtained above are valid for an ideal system with no disorder.  At small wave vectors, a gap opens due to disorder, resulting in a pinning mode that has been studied extensively and is the topic of the subsequent section. As noted in Ref. \onlinecite{Chen2}, it is in principle possible to obtain information about the small $q$ dispersion by comparing the pinning mode resonances measured from nearly identical samples that have different metal film coplanar waveguide (CPW) slot widths $w$ (See any of [\onlinecite{Zhu1,Chen,Chen2,Lewis,Samban}] for experimental details.)
 Measurement of the dispersion at large $q$ is complicated by the fact that light ideally couples only to excitations with very small wave vectors. A possible method for accessing larger momenta through light scattering would be to use gratings or to exploit piezoelectric coupling to surface phonons to define a wave length, which has proven useful in the study of collective mode dispersion of the liquid states of composite fermions.\cite{Kukushkin} The extremal points of the dispersion may also provide a signature in inelastic light scattering, in the presence of disorder, because of the singularity in the density of states at the corresponding energies.\cite{Pinczuk}
 

\subsection{Frequency of the pinning mode}

A number of theoretical studies have considered the electromagnetic response of a pinned Winger solid in a magnetic field,\cite{Fukuyama,Fertig,Fogler,Chitra} and find that the cyclotron resonance is shifted by an amount related to the pinning frequency, and at the same time also broadened. The actual frequencies of the pinning mode depend on the details of the potential and are difficult to predict in a quantitative manner.  We use below the model of Chitra {\em et al.}\cite{Chitra}, where they use a Gaussian variational method to study the elastic Hamiltonian of a Wigner crystal to deduce its properties in the presence of disorder.  

We first address the relation between the pinning mode frequency of the CFWC at an effective filling $\nu^*$ to that of the electron WC at the same filling; we believe that such a relationship might be more stable against the nature of the disorder than the actual pinning frequency. According to Chitra et. al.'s model\cite{Chitra} for the pinning frequencies of the WC, different disorder length scales produce significantly different $B$ dependence for the the pinning frequencies. Their approximate expression for pinning mode frequencies is given by:
\begin{equation}
 \omega^0_p=\frac{\Sigma}{\rho_m\omega_c}
\end{equation}
where $\Sigma=c(2\pi^2)^{-1/6}R_a^{-2}(\frac{a}{\xi_0})^6$, $c \propto e^2\rho^{3/2}$ is the classical shear modulus for the electron WC in units of energy density,
$R_a=\frac{ca^2}{\rho\sqrt{\Delta}}$, $a$ is the lattice spacing, $\Delta$ is a measure of the electrostatic coupling of the particles in the lattice to disorder (and is proportional to the square of the charge of the particle in the lattice), $\xi_0^2=$max[$r_f^2,l^2$], $\rho_m=m/\pi a^2$ is the mass density, $\rho=(\pi a^2)^{-1}$ is the particle density and $r_f$ is the disorder correlation length. The pinning frequencies have a different B-field dependence depending on which of $r_f$ or $l$ is larger. Putting everything together, we find
\begin{equation}
\omega_p^0\sim {a^4 \rho^2 \Delta\over c m\omega_c \xi_0^6}
\end{equation}
Noting that $a$, $\rho$, $m\omega_c$, and $\xi_0$ are identical for electrons near $\nu=n$ and composite fermions near $\nu^*=n$ (The quantity $m\omega_c$ is the same because it is independent of the mass, and $\xi_0$ is the same because the magnetic length and the disorder correlation lengths are unchanged.), the ratio of the pinning mode frequency of composite fermions and electrons, $\omega^{0*}_p$ and $\omega^0_p$, respectively, is given by
\begin{equation}
{\omega^{0*}_p \over \omega^0_p}={c\Delta^* \over c^* \Delta}
\end{equation}
Furthermore, $c\Delta^*$ and $c^*\Delta$ are approximately equal in the classical limit, because both $c$ and $\Delta$ are proportional to $e^2$ whereas $c^*$ and $\Delta^*$ are proportional to $e^{*2}$. Therefore, if we take the coupling to the disorder potential to be the same, the pinning frequencies of the electron WC at $\nu$ and the CFWC at $\nu^*$ are equal in the classical limit, and likely of the same order in the region where the system is not too far from the classical region. While clearly an approximation, this result is consistent with the experimental finding of Zhu {\em et al.},\cite{Zhu1} and, we believe, explains why the pinning frequencies of the electron and CFWCs have similar magnitude in a range of parameters. 

We next come to the $B$-field dependence of the pinning frequency. There is an interesting difference in $B$-field dependence of a type-I and type-II WC. For a type-I crystal, the $B$-field dependence of $\omega_p^0$ depends on 
the relative magnitudes of $l$ and $r_f$. 
The choice $\xi_0=r_f$ gives $\omega_p^0 \propto 1/B$, whereas $\xi_0=l$ produces $\omega_p^0 \propto B^2$. This difference in behavior is easy to identify experimentally and 
the $B$-field dependence of $\omega_p^0$ thus provides evidence for the relevant length scale for disorder. The $B$-field dependence of the type-II WC is more complicated, because the $B$ dependence of the CF quasiparticle/quasihole particle density $\rho^*$ of the type-II WC must be taken into account when examining the $B$-field dependence of $\omega_p^0$. This leads to a simplification for the $B$ dependence of the pinning frequency of the CFWC, because the density variation with $B$ dominates.  Let us consider, for specificity, filling factors in the vicinity of  $\nu=1/3,2/5,3/7$, where $\nu^*=n=\rho\phi_o/B^*$ and $B^*=B(1-2pn)$. As we move slightly away from the rational fillings, we get:
\begin{equation}
 \nu^*=n\pm\bar{\nu}^*=\frac{\rho\phi_0}{B^*\pm\delta B}\approx \frac{\rho\phi_0}{B^*}(1 \pm \delta B/B^*)=\frac{\phi_0}{B^*}(\rho\pm \rho^*)
\end{equation}
where $\rho^*=\rho\delta B/B^*$ (which is equivalent to the relation $\rho^*=\bar{\nu}^*\rho/\nu^*$). 
For the type-I WC we have (using the classical shear modulus) $\omega_p^0 \propto \rho^{-3/2}$, where $\rho$ is independent of $B$ and fixed once the sample is created. For the type-II crystal, when $\bar{\nu}^* \leq .05$, we have $\omega_p^0 \propto (\rho^*)^{-3/2} \propto (\delta B)^{-3/2}$. The dependence of $\omega_p^0$ on $\delta B$ dominates the $B$-field dependence of the pinning mode, regardless of the choice for the disorder length. This leads to the prediction that pinning frequency of the CFWC always decreases as one moves away from the rational fillings, except possibly 
in the vicinity of the melting transition where the shear modulus becomes small. This is consistent with the experimental findings\cite{Zhu1}. Similarly, the $B$-field dependence of the pinning modes of the type-II crystals observed around integer fillings will also be dominated by $\delta B$ and the frequency of the pinning mode will decrease as one moves away from integer fillings, again consistent with the experimental observations\cite{Chen,Lewis}. The $B$-field dependence of the pinning mode thus does not shed qualitative light on the nature of disorder for type-II WCs.

\subsection{Melting Temperature}

\begin{center}
\begin{table}[htb*]
\label{EDCFDDimTable}
\begin{tabular}{|c | c| }\hline 
$\nu^*$ & $T_{\rm M}$ \\ \hline
0.9 & 90 mK \\ \hline
1.06 & 25 mK \\ \hline
1.92 & 10 mK \\ \hline
2.08 & 13 mK \\ \hline
2.94 & 2 mK \\ \hline
3.08 & 72 mK \\
\hline
\end{tabular}
\caption{The peak melting temperatures for CFWC below and above $\nu^*=1$, 2, and 3, which correspond to electrons filling factors in the vicinity of $\nu=1/3$, 2/5, and 3/7. We assume a density of $1.1 \times 10^{11}$ cm$^{-2}$ and parameters appropriate for GaAs.}
\end{table}
\end{center}

\begin{figure}[ht!]
\includegraphics[width=0.45\textwidth]{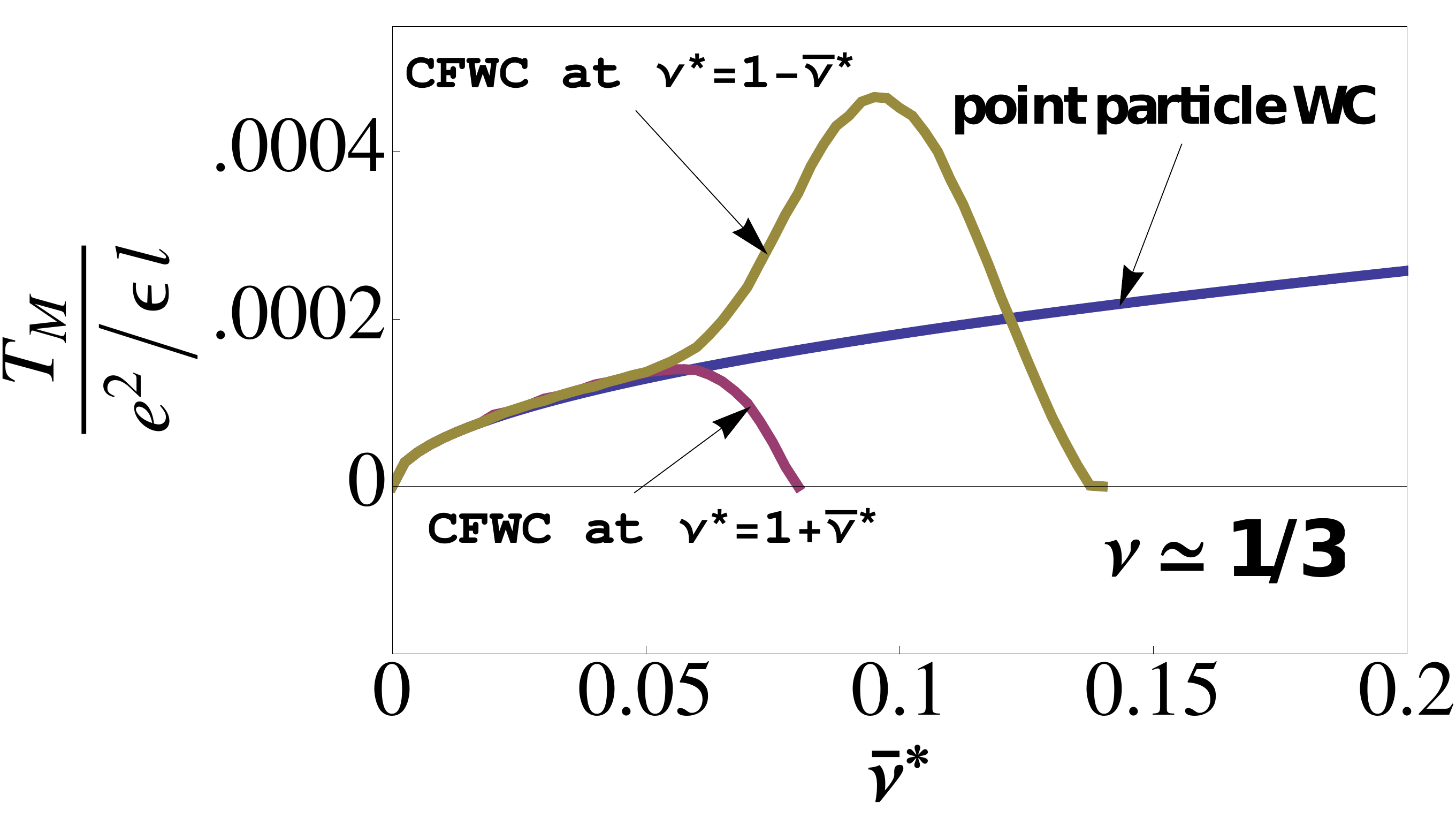}
\includegraphics[width=0.45\textwidth]{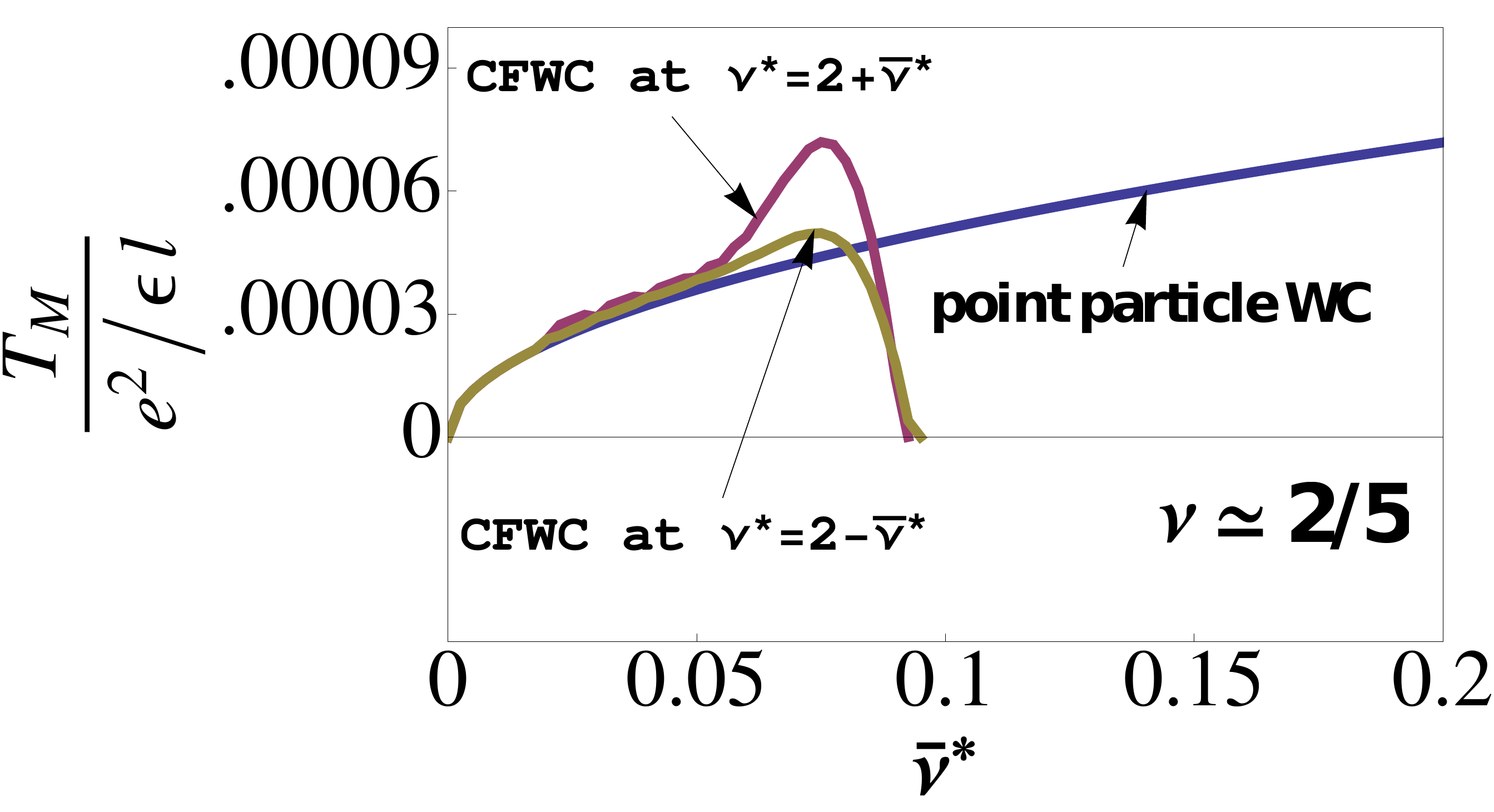}
\includegraphics[width=0.45\textwidth]{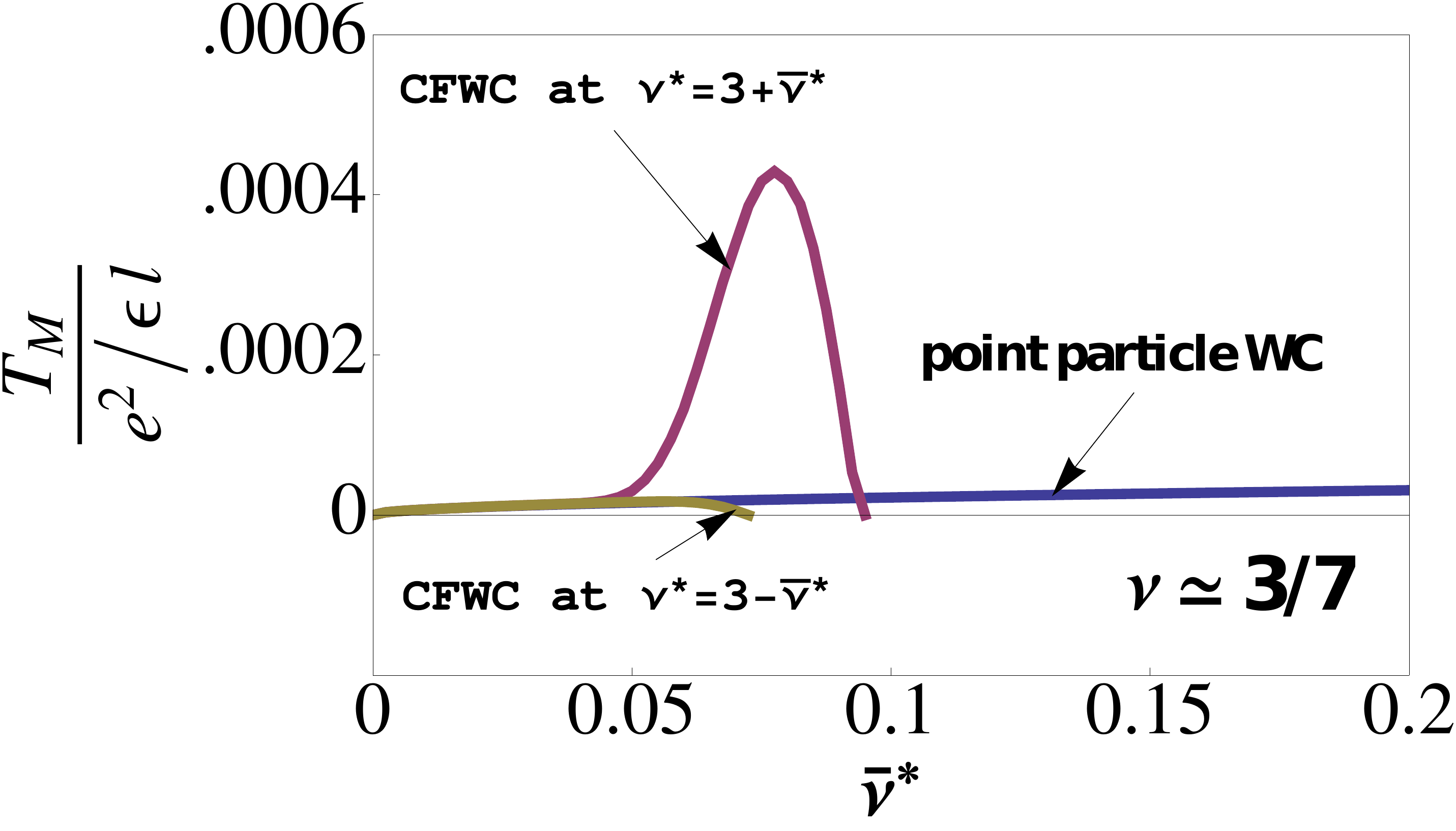}
\caption{Melting temperature $T_{\rm M}$ as a function of the partial filling factor $\bar{\nu}^*=|n-\nu^*|$ where $\nu^*$ is the CF filling factor. The melting temperature is quoted in units of $e^2/\epsilon l$ (with $k_B=1$). The melting temperature of point particles of charge $e^*$ is also included for reference.}\label{MT}
\end{figure}

Another question of interest is the melting temperature of the CFWC. As for two dimensional crystals, it is natural to expect that the melting is described by the Kosertlitz-Thouless mechanism\cite{Koster,Thouless}, with melting temperature given by:
\begin{equation}
T_{\rm M}=(2\pi\sqrt{3})^{-1}C_t
\end{equation}
In Fig.~\ref{MT} we plot the melting temperature of the CFWCs near $\nu=1/3,2/5$ and $3/7$ in units of $e^2/\epsilon l\approx 50 \sqrt{B[T]}\;\text{K}\approx 3.32\times 10^{-4}\sqrt{\rho/\nu}$ K for GaAs-AlGaAs heterostructures, with magnetic field quoted in Tesla, $\rho$ in cm$^{-2}$, and the dielectric function taken to be $\epsilon=12.6$. For a typical density of $\rho=1.1\times 10^{11}$ cm$^{-2}$ at the fillings $\nu=1/3,2/5,3/7$, $e^2/\epsilon l$ corresponds to the temperatures $190\text{K},174\text{K}\text{ and }168\text{K}$, respectively. We also include as reference a plot of the melting temperature of the WC of point particles with charge $e^*$. For illustration, we give in Table I the peak melting temperatures along with the filling factors where they occur. 

Several interesting features are worth noting. The melting temperature at 1/3 is in the same ball park as the experimental temperature of 50 mK where data in Zhu {\emph et. al.} was collected. However, taken at face value, our results would imply a CFWC at the low filling factor side of 1/3 but not at filling factors higher than 1/3. The reason for the discrepancy is unclear at this stage, and may have to do either with some simplifying assumptions of the model (such as neglect or disorder, finite width corrections, or $\Lambda$ level mixing), or with slight intrinsic inaccuracies in the estimation of the effective CF interaction. We also note that the temperature estimated here is when infinitely long range crystalline order is lost; short range crystalline order will possibly persist to higher temperatures. The CFWC melting temperature decreases as we go to 2/5 and 3/7, as expected, making their observation even more challenging; the reason for anomalously high $T_{\rm M}$ for the CF-hole WC at 3/7 is not understood, although it must have to do with the form of the effective interaction. An interesting feature is a significant asymmetry between the behavior of CFWCs above and below $\nu^*=$ integer, as expected from rather general considerations, given that the interaction between CF particles is different from that between CF holes. (Similar asummetry is to be expected for electron WCs above and below integer fillings.) The experimental results appear to suggest a smaller asymmetry than theory; we believe that disorder can possibly wash out the asymmetry. 

Also, note that at low fillings, the melting temperature for the type-II WC near $\nu=n$ is $(2n+1)^2$ larger than the corresponding CFWC near $\nu^*=n$ (since $c/c^*=(2n+1)^2$.) The significantly larger melting temperature of the type-II WC near $\nu=n$  explains the relative ease with which its pinning mode resonance was detected as compared to the resonance near $\nu=1/3$.

\section{Acknowledgments}

We are grateful to the National Science Foundation by partial support under grant no. DMR-1005536, and to the Penn State High-Performance Computing Cluster for computer time.

\appendix

\section{Effective model for Maki-Zotos Hartree-Fock crystal}
\label{MZ}

As indicated in the text, our calculations are based on the Maki-Zotos crystal wave function constructed by placing a single particle wave packet at each lattice site $\vec{R}_j$ of a hexagonal (or triangular) lattice and antisymmetrizing the product.
In this section we outline the derivation of the effective interaction for the CFWC, given in Eq. \!(\ref{V2}). It is not possible to perform this calculation exactly, and one must resort to approximations. We restrict ourselves here to including only the two-body direct and exchange terms since these make the greatest contribution to the energy.\cite{MZ} 

The expression for the denominator of Eq. (\ref{VMZ}) is already given to us by [\onlinecite{MZ}]:
\begin{eqnarray}
 <\! \Psi_{\{\vec{\scriptstyle{R}}_j\}}(\{\vec{r}_i\})|\Psi_{\{\vec{\scriptstyle{R}}_j\}}(\{\vec{r}_i\})\!> & = & \nonumber\\
 1-\frac{1}{2}\sum_{i\neq j}S_{ij}+\cdots&&
\end{eqnarray}
where 
\begin{eqnarray}
S_{ij}\hspace{-1cm} &=& \hspace{-1cm}\int d^2r_1d^2r_2 \psi^*_{\vec{\scriptstyle{R}}_i}(\vec{r}_1) \psi^*_{\vec{\scriptstyle{R}}_j}(\vec{r}_2)\psi_{\vec{\scriptstyle{R}}_j}(\vec{r}_1)\psi_{\vec{\scriptstyle{R}}_i}(\vec{r}_2)\nonumber\\ 
& \hspace{1.15cm}= e^{-\frac{1}{2}R_{ij}^2}. &
\end{eqnarray}
and $R_{ij}=|\vec{R}_i-\vec{R}_j|$. Following appendix A of [\onlinecite{MZ}]:
\begin{eqnarray}
&&<\Psi_{\{\vec{\scriptstyle{R}}_j\}}(\{\vec{r}_i\})|\sum_{i\neq j}V^{\rm eff}(r_{ij})|\Psi_{\{\vec{\scriptstyle{R}}_j\}}(\{\vec{r}_i\})>\nonumber \\
&&\hspace{2cm} =\frac{1}{2}\sum_{i\neq j} V_2(\vec{R}_i,\vec{R}_j)+\cdots
\end{eqnarray}
where
\begin{widetext}
\begin{eqnarray}\label{I1I2}
 V_2(\vec{R}_i,\vec{R}_j) & = & \int d^2r_1d^2r_2 V^{\rm eff}(r_{ij})(|\psi_{\vec{\scriptstyle{R}}_i}(\vec{r}_1)|^2|\psi_{\vec{\scriptstyle{R}}_j}(\vec{r}_2)|^2-\psi^*_{\vec{\scriptstyle{R}}_i}(\vec{r}_1)\psi^*_{\vec{\scriptstyle{R}}_j}(\vec{r}_2)\psi_{\vec{\scriptstyle{R}}_i}(\vec{r}_2)\psi_{\vec{\scriptstyle{R}}_j}(\vec{r}_1)) \nonumber\\
 & = & \frac{1}{(2\pi)^2}\int d^2r_1d^2r_2 V^{\rm eff}(r_{ij}) \big{(}e^{-\frac{1}{2}\left[(\vec{\scriptstyle{r}}_1-\vec{\scriptstyle{R}}_i)^2+(\vec{\scriptstyle{r}}_2-\vec{\scriptstyle{R}}_j)^2\right]}- \nonumber\\
 & & e^{-\frac{1}{4} \left[(\vec{\scriptstyle{r}}_1-\vec{\scriptstyle{R}}_i)^2+(\vec{\scriptstyle{r}}_2-\vec{\scriptstyle{R}}_j)^2+(\vec{\scriptstyle{r}}_1-\vec{\scriptstyle{R}}_j)^2+(\vec{\scriptstyle{r}}_2-\vec{\scriptstyle{R}}_i)^2+2i\left[(x_1-x_2)(Y_i-Y_j)-(y_1-y_2)(X_i-X_j)\right]\right]}\big{)}
\end{eqnarray}
\end{widetext}
Transforming to the center of mass and relative coordinates, the first (direct) term is given by:
\begin{equation}
 I_{1,ij}=\frac{1}{2}e^{-\frac{1}{4}R_{ij}^2}\int rdrV^{\rm eff}(r)e^{-\frac{1}{4}r^2}I_0(\frac{1}{2}R_{ij}r)
\end{equation}
and the second (exchange) term is given by:
\begin{equation}
 I_{2,ij}=\frac{1}{2}e^{-\frac{1}{4}R_{ij}^2}\int rdrV^{\rm eff}(r)e^{-\frac{1}{4}r^2}J_0(\frac{1}{2}R_{ij}r)
\end{equation}
Here, $J_0$ and $I_0$ are the Bessel function and the modified Bessel function of the first kind. The nearest neighbor term $S_{12}$ is of the order $10^{-11}$ for $\nu=.15$. Because we are not concerned with properties of the system when the filling factor is much larger than this, the sum $\sum_{i\neq j}S_{ij}$ is negligible for our purposes. Substituting our results for $I_{1,ij}$ and $I_{2,ij}$ into Eq. (\ref{VMZ}) gives:
\begin{widetext}
\begin{equation}
\frac{<\!\Psi_{\{l\vec{\scriptstyle{R}}_j\}}(\{\vec{r}_i\})|\sum_{i\neq j}V^{\rm eff}(R_{ij})|\Psi_{\{\vec{\scriptstyle{R}}_j\}}(\{\vec{r}_i\})>}    {<\!\Psi_{\{\vec{\scriptstyle{R}}_j\}}(\{\vec{r}_i\})|\Psi_{\{\vec{\scriptstyle{R}}_j\}}(\{\vec{r}_i\})\!>}  \cong \frac{\left[\sum_{i\neq j}I_{1,ij}-I_{2,ij}\right]}{1-\frac{1}{2}\sum_{i\neq j}S_{ij}}  \cong  \sum_{i\neq j}\left[\frac{I_{1,ij}-I_{2,ij}}{1-S_{ij}}\right]
\end{equation}
\begin{equation}
\frac{I_{1,ij}-I_{2,ij}}{1-S_{ij}} = \frac{\frac{1}{2}e^{-\frac{1}{4}R_{ij}^2}\int rdrV^{\rm eff}(r)e^{-\frac{1}{4}r^2}\left(I_0(\frac{1}{2}R_{ij}r)-J_0(\frac{1}{2}R_{ij}r)\right)}{1-e^{-\frac{1}{2}R_{ij}^2}}=\frac{\int rdrV^{\rm eff}(r)e^{-\frac{1}{4}r^2}\left(I_0(\frac{1}{2}R_{ij}r)-J_0(\frac{1}{2}R_{ij}r)\right)}{4\sinh\left(R_{ij}^2/4\right)}
\end{equation}
\end{widetext}
The substitution $1/r$ for $V^{\rm eff}(r)$ reproduces the Maki-Zotos result. For the effective interaction between the composite fermions, the integrals can be performed analytically, but the expressions are quite cumbersome and are most conveniently dealt with in a computer program such as Mathematica.

\section{Dynamical matrix}
\label{DM}

In this section we describe our method for evaluating $\Phi_{\alpha\beta}$, Eq. (\ref{phiab}). We divide $\Phi_{\alpha\beta}$ into two parts, the $k$ independent part and the $k$ dependent part: 
\begin{equation}\label{phiab2}
 \Phi_{\alpha\beta}(\!\vec{k}\!)=\sum_{\vec{R}}\frac{\partial^2V\!(R)}{\partial{R_{\alpha}R_{\beta}}}-\sum_{\vec{R}}\frac{\partial^2V\!(R)}{\partial{R_{\alpha}R_{\beta}}}\cos({\vec{k}\!\cdot\!\vec{R}})
\end{equation}
It is necessary to find a way to compute these sums in a manner so that the error 
can be estimated in a reliable manner, which is especially important for small $k$, where $\Phi_{\alpha\beta}(\!\vec{k}\!)$ can be quite small.
Since both sums are absolutely convergent, we are free to rearrange the terms to produce best convergence.

\begin{figure}[h]
\includegraphics[width=0.45\textwidth]{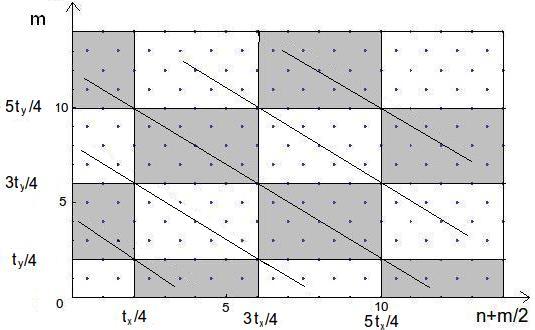}
\caption{
A portion of the upper right hand quadrant of points used in the summation 
of Eq.~(\ref{kdepsum}). The points on the graph mark the locations $(n+m/2,m)$, where $n,m$ are positive integers that occur in the expression Eq.~(\ref{R}). The quantities $t_x\text{ and }t_y$ are defined in Appendix~\ref{DM}.
}\label{kxky}
\end{figure}

For specificity, we describe our method for evaluating $\Phi_{xx}$; extending this discussion to $\Phi_{xy}$ and $\Phi_{yy}$ is straightforward. We begin with the $k$ dependent term in Eq. (\ref{phiab2}). Our goal is to utilize the periodicity of the $\cos(\vec{k}\!\cdot\!\vec{R})$ terms to produce a rapidly convergent alternating series. 
Because of inversion symmetry, we have
\begin{equation}
 \sum_{\vec{R}}\cos({\vec{k}\!\cdot\!\vec{R}})\frac{\partial^2V\!(R)}{\partial{R_x}^2}=\sum_{\vec{R}}\cos(k_xR_x)\cos(k_yR_y)\frac{\partial^2V\!(R)}{\partial{R_x}^2}
\end{equation}
where the sum is over the lattice sites $\vec{R}=a(n+\frac{m}{2},\frac{\sqrt{3}}{2}m)$. Using $a=\sqrt{\frac{4\pi}{\sqrt{3}\bar{\nu}^*}}$ (with $l^*$ as the unit of length), we rewrite this as:
\begin{eqnarray}\label{kdepsum}
&&\sum_{\vec{R}}\cos({\vec{k}\!\cdot\!\vec{R}})\frac{\partial^2V\!(R)}{\partial{R_x}^2}\nonumber \\
&&=\sum_{n,m}\frac{\partial^2V\!(R)}{\partial{R_x^2}}\cos\frac{2\pi(n+m/2)}{t_x}\cos\frac{2\pi m}{t_y}
\end{eqnarray}
where $n,m$ are integers, $\frac{2\pi}{t_x}=ak_x$ and $\frac{2\pi}{t_y}=a\frac{\sqrt{3}}{2}k_y$. We wish to rearrange the sum on the right hand side of Eq. ~\ref{kdepsum} into an alternating series so that it converges more quickly. 
To this end, we divide the integration region into successive light and dark blocks as shown in Fig.~\ref{kxky}, where each point  represents a lattice vector $\vec{R}$. 
The regions are chosen in a manner that $\cos\!\left(\frac{2\pi(n+m/2)}{t_x}\!\right)\cos\!\left(\frac{2\pi m}{t_y}\!\right)$ at any point in the light (dark) regions gives a positive (negative) contribution.
Individual positive or negative terms in the alternating series are constructed by summing over all light or dark blocks along the ``diagonal" lines shown in Fig.~\ref{kxky}. 
We then apply the Euler transformation\cite{Davis} to the resulting alternating series, which produces a rapidly converging sum.

\begin{figure}[h]
\includegraphics[width=0.3\textwidth]{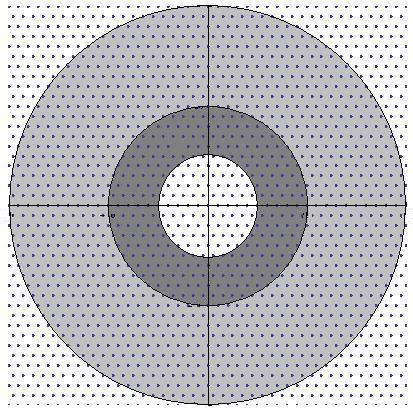}
\caption{Concentric annuli illustrate how the sum in Eq.~(\ref{gensumm}) is organized. The points on the graph have the same locations as in Fig.~\ref{kxky}.
}\label{nkxky}
\end{figure} 

We organize the non $k$-dependent sum in a different manner.  We divide the entire hexagonal lattice into successive annuli (labeled by the index $n$) as in Fig.~\ref{nkxky}, where, beyond certain distance, the inner radius of each successive annulus is twice that of the previous annulus (e.g. the radii of 5, 10, 20 in Fig.~\ref{nkxky}). The partial sum over the points in the $n$th annulus is labeled $s_n$.
The behavior $\frac{\partial^2V\!(R)}{\partial{R_x^2}}\propto\frac{1}{R3}$ at long distances ensures that 
the sum $s_n$ decreases by approximately a factor of two for each successive annulus for large $n$. 
Defining $s_{n+1}/s_n=1/(2+\epsilon_{n+1})$, with $|\epsilon_{n+1}|<|\epsilon_n|$, the sum can be expressed exactly as:
\begin{equation}\label{gensumm}
S=\sum_{\vec{R}}\frac{\partial^2V\!(R)}{\partial{R_x^2}}=s_0+s_1+\cdots+s_n\sum_{i=0}^{\infty}(2+\epsilon_{n+i})^{-i}
\end{equation}
Because we do not know the proper $\epsilon_n$ {\em a priori}, we resort to approximating the remaining sum as a power series by setting $\epsilon_{n+i}=\epsilon_n=0$. The sum is then given by: 
\begin{equation}\label{gensum}
S'=\sum_{\vec{R}}\frac{\partial^2V\!(R)}{\partial{R_x^2}}=s_0+s_1+\cdots+s_n\sum_{i=0}^{\infty}2^{-i}=s_0+s_1+\cdots+2s_n
\end{equation}
To estimate the error in this approximation, we define $S^{\pm}=s_0+s_1+\cdots+s_n\sum_{i=0}^{\infty}(2\mp |\epsilon_{n}|)^{-i}$ so that  $S^-\leq S \leq S^+$. With some algebra, 
and assuming $\epsilon_n\ll1$, this inequality becomes:
\begin{equation}
 s_n(2-|\epsilon_n|) \leq S-s_o-s_1-\cdots - s_{n-1} \leq s_n(2+|\epsilon_n|),
 \end{equation}
which establishes the error to be bounded by $|S-S'|\leq s_n|\epsilon_n|=|2s_n-s_{n-1}|$ 
In our numerical calculations, $\epsilon_n$ becomes rapidly smaller as we increase $n$, allowing us to calculate the sum to a very high degree of accuracy. We implement the sum using the epsilon algorithm given in Ref.~[\onlinecite{Davis}].

\pagebreak


\begin{thebibliography}{99}
\bibitem{commpseudo} The pseudopotentials in the lowest (or any given) LL completely characterize the interaction, and all different choices of interaction that give the same pseudopotential are equivalent.
\end{thebibliography}
\end{document}